\documentclass[prb,showpacs,superscriptaddress,nofootinbib,hypertext,natbib,amsfonts,amssymb,amsmath, twocolumn,aps]{revtex4}
\usepackage{natbib}
\usepackage{graphicx}
\usepackage{graphics}
\usepackage{dsfont}

 \newcommand{\dit}{-0pt}
 
\usepackage{color}

\renewcommand{\url}[1]{}

\allowdisplaybreaks[4]

\begin{document}

\title{Completeness and Bethe root distribution of the spin-$\frac12$ Heisenberg chain with arbitrary boundary fields}

\author{Yuzhu Jiang}
\affiliation{State Key Laboratory of Magnetic Resonance and Atomic
and Molecular Physics, Wuhan Institute of Physics and Mathematics,
Chinese Academy of Sciences, Wuhan 430071, China}
\author{Shuai Cui}
\affiliation{Beijing National Laboratory for Condensed Matter
Physics, Institute of Physics, Chinese Academy of Sciences, Beijing
100190, China}
\author{Junpeng Cao}
\affiliation{Beijing National Laboratory for Condensed Matter
Physics, Institute of Physics, Chinese Academy of Sciences, Beijing
100190, China}
\author{Wen-Li Yang}
\affiliation{Institute of Modern Physics, Northwest University,
Xi¡¯an 710069, China}
\author{Yupeng Wang\footnote{Corresponding author: yupeng@iphy.ac.cn}}
\affiliation{Beijing National Laboratory for Condensed Matter
Physics, Institute of Physics, Chinese Academy of Sciences, Beijing
100190, China}

\begin{abstract}
Recently, the $XXX$ spin chain with arbitrary boundary fields was
successfully solved\cite{cysw2} via the off-diagonal Bethe ansatz
method\cite{cysw1}. The correctness and the completeness of this
solution were numerically verified by Nepomechie for one choice of
the parameterizations.\cite{nepo} In this paper, we discuss further
the completeness of another parameterization of the Bethe ansatz
equations and its reduction to the parallel boundary field case. The
numerical results show that when the relative angle between the
boundary fields turns to zero, both the $T-Q$ relations and the
Bethe ansatz equations are reduced to the ones obtained by the
conventional Bethe ansatz methods. This allows us to establish a
one-to-one correspondence between the Bethe roots of the unparallel
boundary field case and those of the parallel boundary field case.
In the thermodynamic limit $N\to \infty$, those two sets of Bethe
roots tend to the same and the contribution of the relative angle to
the energy is in the order of $N^{-1}$.
\end{abstract}


\maketitle

\section{Introduction}

The Heisenberg spin chain model plays a very important role in
condensed matter physics since it provides a benchmark for
understanding the magnetic properties in one dimension. This model
was firstly solved by Bethe with which the so-called Bethe ansatz
method was invented.\cite{bethe} The open boundary problem of this
model was initially studied by Gaudin.\cite{gaudin} Thereafter the
model with parallel
boundary fields was successfully solved in Ref. 6. 
An important achievement for the integrable models with open
boundaries is the reflection equation proposed in Ref. 7
which induces the important result: The spin chain with arbitrary
unparallel boundary fields is also integrable! This problem has
attracted a lot of attentions \cite{npb03,Nep04, Yan04, fra1,fra2,
nico1,nico2,nico3} not only for its mathematical structure but also
for its relevance to the stochastic process and the spin current
physics. The density matrix renormalization group (DMRG) analysis
showed that the spin-spin correlation in this model may show a
spiral behavior and the unparallel boundary fields may induce a
``spin voltage". \cite{zhuo} Though the integrablity has been
demonstrated for over two decades, the coordinate Bethe ansatz and
the algebraic Bethe ansatz encountered a big problem for approaching
such a kind of models because those methods strongly rely on the
existence of a local vacuum which these model do not possess as the
$U(1)$ symmetry is broken by the unparallel boundaries. Until very
recently, the model was solved\cite{cysw2} via the off-diagonal
Bethe ansatz method\cite{cysw1,cysw3}. This method overcomes the
obstacle of the lack of a reference state and allows to derive the
eigenvalues of the transfer matrix only based on some operator
product identities. It is shown that the the form of the Bethe
ansatz equations (BAEs) is quite different from that of the parallel
boundary case and may be parameterized in a variety of different
ways. The completeness of the solutions was numerically checked by
Nepomechie for a special choice of the parameterizations and the
Bethe root distribution in the ground state is also
discussed.\cite{nepo}

In this paper, we examine further the completeness of another
parameterization of the BAEs numerically up to lattice number $N=6$
by comparison of  numerically solutions of these equations and the
results from the exact diagonalization. In addition, we also
consider the reduction of the BAEs to those of the parallel boundary
case when the relative angle between the two boundary fields
$\alpha\to0$. Such a reduction process allows us to establish the
one-to-one correspondence between the Bethe roots of the two cases.
With this correspondence a perturbation theory can be used to derive
the physical effect of the boundary fields.

The paper is organized as follows. In Sec.\ref{SecME}, we give a
brief description of the model and the BAEs. The numerical check for
the completeness of the BAEs and the distribution of Bethe roots is
discussed in Sec.\ref{SecCBR} with $N=3,5$ and $N=4$ as two examples
representing the odd $N$ and even $N$ cases, respectively. The
reduction from the case of the unparallel boundary fields to that of
the parallel boundary fields is given in Sec.\ref{SecRDC}. This
allows us to construct the correspondence between the two sets of
Bethe roots. Sec.\ref{SecTL} is attributed to the effect of the
relative angle between the boundary fields in the thermodynamic
limit $N\to\infty$. The concluding remarks are given in
Sec.\ref{SecC}.

\section{Model and exact solution}
\label{SecME}

The Hamiltonian of the model reads
\begin{align}
 \label{H}
 H&=\sum_{j=1}^{N-1} \vec \sigma_j \vec \sigma_{j+1}
   +\frac{1}{p} \sigma^z_N + \frac{1}{q}(\sigma^z_1+\xi\sigma^x_1).
\end{align}
where $p,q$ and $\xi$ are arbitrary real constants and $\vec
\sigma_j$ are the Pauli matrices as usual. $1/p$ and
$q_0=q/\sqrt{1+\xi^2}$ represent the strengths of the two boundary
fields, respectively. Notice that we can always choose one boundary
field along $z$-axis and the other in the $x-z$ plane without losing
generality because of the $O(3)$-invariance of the bulk. For
convenience, we define the relative angle between the two boundary
fields as $\alpha=\arctan\xi$ and $\alpha\in(-\pi/2,\pi/2)$.

In Ref. 1 
, a general expression of the eigenvalue
$\Lambda(u)$ of the transfer matrix was derived based on the
operator product identities of the transfer matrix. In this paper,
we choose the following extended $T-Q$ ansatz of $\Lambda(u)$
\begin{align}
 \label{TQ}
 \varLambda(u)
 =\bar a(u)\frac{Q_1(u-1)}{Q_2(u)}
 +\bar d(u)\frac{Q_2(u+1)}{Q_1(u)}
 +\frac{\bar c(u)\bar a(u) \bar d(u)}{Q_1(u)Q_2(u)},
\end{align}
with
\begin{align}
 &
 \bar a(u)=\sqrt{1+\xi^2}\frac{u+1}{u+\frac12}
 (u+p)(u+q_0)(u+1)^{2N},\nonumber\\
 &
 \bar d(u)=\sqrt{1+\xi^2}\frac{u}{u+\frac12}
 (u+1-p)(u+1-q_0)u^{2N},\nonumber\\
 &
 \bar c(u)=-\frac{\xi_\alpha}{2\sqrt{1+\xi^2}}
 \frac{(2u+1)^{2}}{u+1-p}
 \frac{[u(u+1)]^{-\beta}}{u+1-q_0}
 \frac{1}{u+p}\frac1{u+q_0},\nonumber\\
 &
 Q_1(u)=\prod_{j=1}^M\Big(u-{\rm i}\mu_j+\frac12\Big),
 Q_2(u)=\prod_{j=1}^M\Big(u+{\rm i}\mu_j+\frac12\Big),
 \nonumber
\end{align}
where $\xi_\alpha=4\sin^2(\alpha/2)$, $M\geq N$ is an integer and
$\mu_j$'s are the Bethe roots. $\varLambda(u)$ possesses the
following properties: (1) It is a polynomial of degree $2N+2$; (2)
It satisfies the the crossing symmetry relation
$\varLambda(u)=\varLambda(-u-1)$. Notice that $\bar a(-u-1)=\bar
b(u)$, $\bar c(-u-1)=\bar c(u)$ and $Q_1(-u-1)=(-1)^MQ_2(u)$ protect
the crossing symmetry automatically. Since $\Lambda(u)$ is a
polynomial, the regularity induces the following BAEs
\begin{align}
 \label{BAE0}
 &\xi_\alpha \Big[
 \Big(\mu_j+\frac{\rm i}2\Big)
 \Big(\mu_j-\frac{\rm i}2\Big)\Big]^\beta \mu_j
 \Big(\mu_j-\frac{\rm i}{2}\Big)^{2N+1}
 =(\mu_j+{\rm i}p')\nonumber\\
 &\times
 (\mu_j+{\rm i}q')
 \prod_{l=1}^M
 (\mu_j+\mu_l)(\mu_j+\mu_l-{\rm i}),j=1,2,\cdots,M,
\end{align}
where $p'=p-1/2$, $q'=q_0-1/2$ and $\beta=M-N$. Notice that the
simplicity of the ``poles" is required to keep the self-consistency
in deriving the BAEs. This requirement gives the following selection
rules for the Bethe roots: $\mu_l\neq\mu_j$ for $l\neq j$ and
$\mu_j+\mu_l \neq 0,{\rm i}$. The integer $\beta$ here should be
nonnegative even number for even $N$  and positive odd number for
odd $N$. In the numerical verifications, we choose $M=N$ and
$\beta=0$ for even $N$ and $M=N+1$ and $\beta=1$ for odd $N$. The
eigenvalue of Hamiltonian (\ref{H}) is given by
\begin{align}
 \label{eig0}
 E=N-1+\frac{1}{p}+\frac{1}{q_0}
  -2\sum_{j=1}^M \frac{1}{{\rm i}\mu_j+1/2}.
\end{align}

\section{Completeness of the Bethe ansatz solutions}
\label{SecCBR} The numerical solutions for Eqs. (\ref{BAE0}) are
undertaken for some given parameters  $p$, $q$ and $\xi$ with small
$N$. The results for $p=-0.6,~ q=-0.3, \xi=1.2$ and $N=3,4,5$ are
listed in TABLE \ref{BAEnums-3}-\ref{BAEnums-5} respectively. The
values of the eigen energy in the tables are calculated from both
the Bethe roots and the exact diagonalization, which coincides
exactly within the computation accuracy.
\begin{table}
\caption{\label{BAEnums-3}
 Bethe roots for $p=-0.6, q=-0.3, \xi=1.2$ $N=3$. elv indicates the energy levels calculated from both the Bethe roots and the exact diagonalization}
\begin{ruledtabular}
\begin{tabular}{cc|rr}
$\mu_{1,2}$ &$\mu_{3,4}$ & $E$ & elv\\
\hline
$\pm 0.40495-0.01585{\rm i}$  & $\pm 0.25771+0.50018{\rm i}$  & $-9.66040$ & 1 \\[\dit]
$\pm 0.36230-0.00003{\rm i}$  & $\pm 3.08784{\rm i}-0.48274{\rm i}$ & $-9.66040$ & 1 \\[\dit]
$\pm 0.16920+0.50000{\rm i}$  & $\pm 1.57781{\rm i}-0.71570{\rm i}$ & $-5.22656$ & 2 \\[\dit]
$\pm 1.57984-0.84026{\rm i}$  & $\pm 1.51070+1.08886{\rm i}$  & $-5.22656$ & 2 \\[\dit]
$\pm 0.14505+0.04771{\rm i}$  & $\pm 0.13230+0.92111{\rm i}$  & $-4.24721$ & 3 \\[\dit]
$\pm 1.27132+0.04315{\rm i}$  & $\pm 1.27426{\rm i}-0.04143{\rm i}$ & $-4.24721$ & 3 \\[\dit]
$\pm 0.33201+0.50005{\rm i}$  & $\pm 1.45102{\rm i}-0.33565{\rm i}$ & $-2.49645$ & 4 \\[\dit]
$\pm 4.11978{\rm i}-0.79915{\rm i}$ & $\pm 1.11460{\rm i}-0.00355{\rm i}$ & $-2.49645$ & 4 \\[\dit]
$\pm 0.88730+0.02945{\rm i}$  & $\pm 0.69758{\rm i}-0.00590{\rm i}$ & $ 2.03247$ & 5 \\[\dit]
$\pm 0.28427+0.00744{\rm i}$  & $\pm 0.29118{\rm i}+0.98321{\rm i}$ & $ 2.03247$ & 5 \\[\dit]
$\pm 0.41166+0.50097{\rm i}$  & $\pm 0.86138{\rm i}-0.16936{\rm i}$ & $ 4.25829$ & 6 \\[\dit]
$\pm 3.58744{\rm i}-0.63520{\rm i}$ & $\pm 0.69229{\rm i}-0.00029{\rm i}$ & $ 4.25829$ & 6 \\[\dit]
$\pm 0.83111-0.04440{\rm i}$  & $\pm 0.20999{\rm i}+0.90204{\rm i}$ & $ 6.60218$ & 7 \\[\dit]
$\pm 1.13291{\rm i}-0.12386{\rm i}$ & $\pm 0.67782{\rm i}-0.01441{\rm i}$ & $ 6.60218$ & 7 \\[\dit]
$\pm 1.09568{\rm i}-0.55876{\rm i}$ & $\pm 0.19858{\rm i}+0.89064{\rm i}$ & $ 8.73767$ & 8 \\[\dit]
$\pm 5.57838{\rm i}-1.52046{\rm i}$ & $\pm 0.20398{\rm
i}+0.89604{\rm i}$ & $ 8.73767$ & 8
\end{tabular}
\end{ruledtabular}
\end{table}

\begin{table}
\caption{\label{BAEnums-4}
 Bethe roots and the eigen energies for $p=-0.6, q=-0.3, \xi=1.2$ $N=4$.}
\begin{ruledtabular}
\begin{tabular}{cc|rr}
$\mu_{1,2}$ &$\mu_{3,4}$ & $E$ & elv\\
\hline
$\pm 1.36767+0.21035{\rm i}$  & $\pm 0.26039+0.00024{\rm i}$  & $-10.76127$ & 1 \\[0pt]
$\pm 0.20995+0.00021{\rm i}$  & $\pm 1.69238{\rm i}-0.28144{\rm i}$ & $-9.28939$ & 2 \\[0pt]
$\pm 0.79468-0.43736{\rm i}$  & $\pm 0.70776+0.46346{\rm i}$  & $-6.64726$ & 3 \\[0pt]
$\pm 0.48700-0.00183{\rm i}$  & $\pm 1.35733{\rm i}-0.19102{\rm i}$ & $-5.95319$ & 4 \\[0pt]
$\pm 1.34713+0.49283{\rm i}$  & $\pm 2.62865{\rm i}-0.77106{\rm i}$ & $-4.04791$ & 5 \\[0pt]
$\pm 0.16855-0.10453{\rm i}$  & $\pm 0.12746+1.06030{\rm i}$  & $-3.22515$ & 6 \\[0pt]
$\pm 1.77115+0.22696{\rm i}$  & $\pm 1.15047{\rm i}-0.04554{\rm i}$ & $-2.08666$ & 7 \\[0pt]
$\pm 3.46475{\rm i}-0.75587{\rm i}$ & $\pm 1.14892{\rm i}-0.04841{\rm i}$ & $-1.23956$ & 8 \\[0pt]
$\pm 0.34771-0.00294{\rm i}$  & $\pm 0.78950{\rm i}-0.09748{\rm i}$ & $-0.28153$ & 9 \\[0pt]
$\pm 0.19568-0.00928{\rm i}$  & $\pm 0.26033{\rm i}+0.95237{\rm i}$ & $ 2.50287$ & 10 \\[0pt]
$\pm 1.56553+0.22572{\rm i}$  & $\pm 0.70522{\rm i}-0.01317{\rm i}$ & $ 4.46454$ & 11 \\[0pt]
$\pm 0.45569+0.01413{\rm i}$  & $\pm 0.21313{\rm i}+0.90518{\rm i}$ & $ 5.39498$ & 12 \\[0pt]
$\pm 2.62317{\rm i}-0.52576{\rm i}$ & $\pm 0.71018{\rm i}-0.01812{\rm i}$ & $ 5.61585$ & 13 \\[0pt]
$\pm 1.37968{\rm i}-0.29819{\rm i}$ & $\pm 0.64194{\rm i}-0.05012{\rm i}$ & $ 7.22925$ & 14 \\[0pt]
$\pm 1.21382-0.10390{\rm i}$  & $\pm 0.20470{\rm i}+0.89675{\rm i}$ & $ 8.55124$ & 15 \\[0pt]
$\pm 7.91636{\rm i}-2.01412{\rm i}$ & $\pm 0.20397{\rm i}+0.89603{\rm i}$ & $ 9.77318$ & 16 \\[0pt]
\end{tabular}
\end{ruledtabular}
\end{table}

\begin{table}
\caption{\label{BAEnums-5}
 Bethe roots and the eigen energies for $p=-0.6, q=-0.3, \xi=1.2$ $N=5$.}
\begin{ruledtabular}
\begin{tabular}{ccc|rr}
$\mu_{1,2}$ &$\mu_{3,4}$ &$\mu_{5,6}$ & $E$ & elv\\
\hline
$\pm 0.52-0.00{\rm i}$  & $\pm 0.20-0.00{\rm i}$  & $\pm 3.12{\rm i}-0.49{\rm i}$  & $-13.1469$ & 1 \\[0pt]
$\pm 1.65-0.80{\rm i}$  & $\pm 1.60+1.09{\rm i}$  & $\pm 0.19-0.00{\rm i}$  & $-10.2172$ & 2 \\[0pt]
$\pm 1.51+0.06{\rm i}$  & $\pm 0.17-0.00{\rm i}$  & $\pm 1.42{\rm i}-0.05{\rm i}$  & $-9.4824$ & 3 \\[0pt]
$\pm 0.14-0.00{\rm i}$  & $\pm 4.25{\rm i}-0.84{\rm i}$  & $\pm 1.14{\rm i}-0.00{\rm i}$  & $-8.0481$ & 4 \\[0pt]
$\pm 1.56-0.81{\rm i}$  & $\pm 1.50+1.06{\rm i}$  & $\pm 0.47+0.00{\rm i}$  & $-7.5446$ & 5 \\[0pt]
$\pm 1.35+0.05{\rm i}$  & $\pm 0.37+0.00{\rm i}$  & $\pm 1.20{\rm i}-0.03{\rm i}$  & $-7.1127$ & 6 \\[0pt]
$\pm 0.21-0.50{\rm i}$  & $\pm 0.21+0.50{\rm i}$  & $\pm 3.65{\rm i}-0.66{\rm i}$  & $-6.3131$ & 7 \\[0pt]
$\pm 0.33+0.00{\rm i}$  & $\pm 4.17{\rm i}-0.82{\rm i}$  & $\pm 1.11{\rm i}-0.00{\rm i}$  & $-6.0858$ & 8 \\[0pt]
$\pm 1.09-0.02{\rm i}$  & $\pm 0.17+0.50{\rm i}$  & $\pm 1.42{\rm i}-0.59{\rm i}$  & $-4.4728$ & 9 \\[0pt]
$\pm 0.58-0.52{\rm i}$  & $\pm 0.57+0.50{\rm i}$  & $\pm 1.21{\rm i}-0.08{\rm i}$  & $-3.8603$ & 10 \\[0pt]
$\pm 0.69-0.00{\rm i}$  & $\pm 4.00{\rm i}-0.76{\rm i}$  & $\pm 1.11{\rm i}-0.00{\rm i}$  & $-3.2223$ & 11 \\[0pt]
$\pm 0.20+0.50{\rm i}$  & $\pm 0.08-0.45{\rm i}$  & $\pm 0.32{\rm i}+1.43{\rm i}$  & $-2.1527$ & 13 \\[0pt]
$\pm 1.11+0.05{\rm i}$  & $\pm 0.26+0.00{\rm i}$  & $\pm 0.70{\rm i}-0.00{\rm i}$  & $-1.6813$ & 14 \\[0pt]
$\pm 1.98+1.21{\rm i}$  & $\pm 1.95-0.76{\rm i}$  & $\pm 1.10{\rm i}-0.00{\rm i}$  & $-1.0893$ & 15 \\[0pt]
$\pm 1.58+0.42{\rm i}$  & $\pm 0.37+0.50{\rm i}$  & $\pm 1.30{\rm i}-0.20{\rm i}$  & $-0.6705$ & 16 \\[0pt]
$\pm 0.43+0.50{\rm i}$  & $\pm 0.23+0.00{\rm i}$  & $\pm 0.89{\rm i}-0.20{\rm i}$  & $-0.3244$ & 17 \\[0pt]
$\pm 6.48{\rm i}-1.52{\rm i}$  & $\pm 2.21{\rm i}-0.05{\rm i}$  & $\pm 1.10{\rm i}-0.00{\rm i}$  & $-0.0889$ & 18 \\[0pt]
$\pm 0.97{\rm i}-0.14{\rm i}$  & $\pm 0.66{\rm i}-0.03{\rm i}$  & $\pm 0.16{\rm i}-0.00{\rm i}$  & $ 1.5688$ & 19 \\[0pt]
$\pm 0.60-0.01{\rm i}$  & $\pm 0.39+0.50{\rm i}$  & $\pm 0.83{\rm i}-0.14{\rm i}$  & $ 2.9018$ & 20 \\[0pt]
$\pm 0.13+0.00{\rm i}$  & $\pm 1.12{\rm i}-0.59{\rm i}$  & $\pm 0.19{\rm i}+0.88{\rm i}$  & $ 3.2817$ & 21 \\[0pt]
$\pm 0.92-0.05{\rm i}$  & $\pm 0.37-0.00{\rm i}$  & $\pm 0.21{\rm i}+0.90{\rm i}$  & $ 3.7343$ & 22 \\[0pt]
$\pm 0.34-0.00{\rm i}$  & $\pm 1.11{\rm i}-0.57{\rm i}$  & $\pm 0.20{\rm i}+0.89{\rm i}$  & $ 5.1951$ & 23 \\[0pt]
$\pm 0.23+0.50{\rm i}$  & $\pm 0.16-0.85{\rm i}$  & $\pm 1.00{\rm i}+1.69{\rm i}$  & $ 5.5261$ & 24 \\[0pt]
$\pm 0.92+0.48{\rm i}$  & $\pm 0.60+0.50{\rm i}$  & $\pm 0.81{\rm i}-0.12{\rm i}$  & $ 6.0442$ & 25 \\[0pt]
$\pm 5.42{\rm i}-1.20{\rm i}$  & $\pm 1.76{\rm i}-0.02{\rm i}$  & $\pm 0.69{\rm i}-0.00{\rm i}$  & $ 6.8680$ & 26 \\[0pt]
$\pm 0.57+0.50{\rm i}$  & $\pm 0.47-0.41{\rm i}$  & $\pm 0.21{\rm i}+0.90{\rm i}$  & $ 7.3435$ & 27 \\[0pt]
$\pm 0.65+0.01{\rm i}$  & $\pm 1.07{\rm i}-0.53{\rm i}$  & $\pm 0.20{\rm i}+0.89{\rm i}$  & $ 7.8245$ & 28 \\[0pt]
$\pm 4.43{\rm i}-0.90{\rm i}$  & $\pm 1.10{\rm i}-0.01{\rm i}$  & $\pm 0.69{\rm i}-0.00{\rm i}$  & $ 8.2802$ & 29 \\[0pt]
$\pm 1.08-0.75{\rm i}$  & $\pm 1.03+0.52{\rm i}$  & $\pm 0.20{\rm i}+0.90{\rm i}$  & $ 9.1338$ & 30 \\[0pt]
$\pm 1.57-0.00{\rm i}$  & $\pm 5.31{\rm i}-1.37{\rm i}$  & $\pm 0.20{\rm i}+0.90{\rm i}$  & $ 10.0097$ & 31 \\[0pt]
$\pm 10.44{\rm i}-2.58{\rm i}$  & $\pm 1.04{\rm i}-0.50{\rm i}$  & $\pm 0.20{\rm i}+0.90{\rm i}$  & $ 10.7940$ & 32 \\[0pt]
\end{tabular}
\end{ruledtabular}
\end{table}
The numerical results for $N=6$ (not listed) also indicate the BAEs
give the complete solutions of the spectrum. For even $N$, we have
$2^N$ set of solutions for the Bethe roots which correspond to the
$2^N$ eigenstates. In addition, the roots in this case possess the
self-opposite-conjugate property, i.e., if $\mu_j$ is a root,
$-\mu_j^*$ also exist in the same set of solution. For odd $N$,
however, one energy level corresponds to two set of Bethe roots,
i.e., if we have one set of Bethe roots, its opposite conjugate
forms another set of solution which correspond to the same
eigenstate. To make the table shorter, we only chose one of the two
sets of the roots in TABLE \ref{BAEnums-5}. Even though we can not
demonstrate the completeness of the solutions from the BAEs, the
numerical results strongly suggest that the solution is complete. In
fact, from the BAEs we learn that there are infinite number of
choices of $\beta$. Different choice only gives different
parameterization but not different solution. In such a sense, we may
believe the solution from the BAEs is complete.

\section{Reduction to the parallel boundary field case}~
\label{SecRDC}

When $\alpha$ tends to zero, the Hamiltonian is reduced to the case
studied in Refs. 6 and 7. 
The BAEs obtained from
the off-diagonal Bethe ansatz should  also be reduced to those of
the parallel boundary field case. For a very small $\alpha$,
$\xi_\alpha =\alpha^2 +O(\alpha^4)$. From Eq.(\ref{BAE0}), we can
see that the left hand side must be also very small. Thus, in the
small $\alpha$ limit, we can divide the $\mu_j$ into two classes
\begin{align}
 \label{reu}
 &\mu^{x}_\pm=\pm x_l+{\rm i}\delta^x_l, &l=1,2,\cdots,M_1;
 \nonumber\\
 &\mu^{y}_\pm=\pm y_l+{\rm i}/2+{\rm i}\delta^y_l, &l=1,2,\cdots,M_2.
\end{align}
where $\mu^{x}_++\mu^{x}_-=2{\rm i}\delta^x_l$,
$\mu^{y}_++\mu^{y}_-=2{\rm i}\delta^y_l$, $\delta^{x,y}$ are small
numbers and $M=2(M_1+M_2)$. Based on this classification of $\mu_j$,
both the $T-Q$ relation and the BAEs can be reduced to those of the
parallel boundary field case. For $\alpha=0$, $\delta^{x,y}=0$.
Submitting these results into Eq.(\ref{TQ}), the extended $T-Q$
relation is reduced to
\begin{align}
 \label{rtq}
 \varLambda&(\mu)
 ={\bar a}(\mu)\frac{\bar Q(\mu-1)}{\bar Q(\mu)}
 +{\bar d}(\mu)\frac{\bar Q(\mu+1)}{\bar Q(\mu)},
\end{align}
where
\begin{align}
\bar Q(u)=\prod_{j=1}^{M_1}
 \Big(\mu-{\rm i}x_j+\frac12\Big)
 \Big(\mu+{\rm i}x_j+\frac12\Big),
\end{align}
which is just the $T-Q$ relation for the parallel boundary field
case.

Now let us check the BAEs. Submitting Eq. (\ref{reu}) into Eq.
(\ref{BAE0}), we get four sets of equations of $\mu^x_\pm$ and
$\mu^y_\pm$ respectively. Keeping the leading order of
$\delta^{x,y}$, we get the following four equations:
\begin{align}
 &\label{BAErx}
 \Big(\frac{x_j-\frac{\rm i}{2}}
 {x_j+\frac{\rm i}{2}}\Big)^{2N+1}
 \hspace{-10pt}=-
 \frac{x_j+{\rm i}p'}{x_j-{\rm i}p'}
 \frac{x_j+{\rm i}q'}{x_j-{\rm i}q'}
 \prod_{l,r}^{M_1}
 \frac{x_j+rx_l-{\rm i}}{x_j+rx_l+{\rm i}},\\
 &\label{BAEry}
 \Big(\frac{y_j+{\rm i}}{y_j+{\rm i}}\Big)^{\beta}
 \frac{y_j-\frac{\rm i}{2}}{y_j+\frac{\rm i}{2}}
 =-
 \frac{y_j+{\rm i}p'}{y_j-{\rm i}p'}
 \frac{y_j+{\rm i}q'}{y_j-{\rm i}q'}
 \prod_{l,r}^{M_2}
 \frac{y_j+ry_l-{\rm i}}{y_j+ry_l+{\rm i}}
 \\
 & \label{BAErdx}
 {\rm i}\delta^x_j =\frac{\xi_\alpha}{4}
 \frac{[(x_j+\frac{\rm i}{2})(x_j-\frac{\rm i}{2})]^\beta}
 {(x_j+{\rm i}p')(x_j+{\rm i}q')}
 \frac{(x_j-\frac{\rm i}{2})^{2N+1}}{\prod'^{M_1}_{l,r}(x_j+rx_l)}
 \nonumber\\
 & \times\prod_{l,r}^{M_1}
 \frac1{x_j+rx_l-{\rm i}}
 \prod_{l,r}^{M_2}\frac1{x_j+ry_l+\frac{\rm i}{2}}\frac1{x_j+ry_l-\frac{\rm i}2},
 \\
 & \label{BAErdy}
 {\rm i}\delta^y_j
 =\frac{\xi_\alpha}{4}
 \frac{[(y_j+\frac{\rm i}{2})(y_j-\frac{\rm i}{2})]^\beta}
 {(y_j+{\rm i}p')(y_j+{\rm i}q')}
 \frac{(y_j-\frac{\rm i}{2})^{2N+1}}{\prod'^{M_2}_{l,r}(y_j+ry_l)}
 \nonumber\\
 & \times\prod_{l,r}^{M_2}\frac1{y_j+ry_l-{\rm i}}
 \prod_{l,r}^{M_1}\frac1{y_j+rx_l+\frac{\rm i}2}\frac1{x_j+ry_l-\frac{\rm i}2},
\end{align}
where $r=\pm$. Notice that in Eqs. (\ref{BAErx}) and (\ref{BAEry})
 $x_j$ and $y_j$ are decoupled from each other. When $\alpha=0$, $\xi_0=0$,
 $\delta^{x,y}=0$,
from the energy expression (\ref{eig0}), we can see that only $x_j$
contribute and $y_j$ are irrelevant. The equations about $x_j$ are
just the BAEs of the parallel boundary field case. \cite{alcraz,skl}
\begin{table*}
\caption{\label{tablered} Correspondence of Bethe roots of the
ground state for $p=0.6$, $q_0=0.6$ and $N=10$.}
\begin{ruledtabular}
\begin{tabular}{c|ccccc}
$\alpha$ & $\mu^x_{+,1}$ & $\mu^x_{+,2}$ & $\mu^x_{+,3}$ & $\mu^x_{+,4}$ & $\mu^x_{+,5}$\\
\hline
$0.0\pi $& $0.06215-0.00000{\rm i}$& $0.13790-0.00000{\rm i}$& $0.23440-0.00000{\rm i}$& $0.36296+0.00000{\rm i}$& $0.56279-0.00000{\rm i}$\\
$0.1\pi $& $0.05834+0.00942{\rm i}$& $0.14116+0.01253{\rm i}$& $0.24263+0.01139{\rm i}$& $0.37557+0.00980{\rm i}$& $0.58306+0.00826{\rm i}$\\
$0.3\pi $& $0.05426+0.02488{\rm i}$& $0.14343+0.03075{\rm i}$& $0.24972+0.03269{\rm i}$& $0.39081+0.03666{\rm i}$& $0.62243+0.04782{\rm i}$\\
$0.4\pi $& $0.05337+0.02967{\rm i}$& $0.14383+0.03605{\rm i}$& $0.25099+0.03927{\rm i}$& $0.39382+0.04611{\rm i}$& $0.63233+0.06678{\rm i}$\\
$0.5\pi $& $0.05278+0.03332{\rm i}$& $0.14408+0.04004{\rm i}$& $0.25180+0.04427{\rm i}$& $0.39569+0.05347{\rm i}$& $0.63838+0.08281{\rm i}$\\
\end{tabular}
\end{ruledtabular}
\end{table*}

For the traditional BAEs, to find the complete solutions we need to
choose $M_1$ from $0$ to $N$. To get a complete reduction, we can
set $\beta\geq N$, for example $M=2N$. The correctness of this
reduction process is also checked numerically for  $p=0.6$,
$q_0=0.6$ and $N=10$. The ground state Bethe root distributions for
$\alpha=0, 0.1\pi, 0.3\pi, 0.4\pi$ and $0.5\pi$ are listed in TABLE
\ref{tablered}. The data listed in the first line are the five $x_l$
obtained from the conventional BAEs; while the other lines are those
for nonzero $\alpha$.  $\mu^{x}_+$ indicate the five $\mu_j$ with
positive real part and the other five are $\mu^x_-=-(\mu^x_+)^*$.
From the numerical results we can see that with the decreasing of
$\alpha$, the Bethe roots tend to the values of $\alpha=0$ case.

\section{The thermodynamic limit}
\label{SecTL}

In the thermodynamic limit $N\to\infty$, the contribution of
$\alpha$ should become smaller and smaller with the increasing $N$.
From Eq. (\ref{BAE0}) we get
\begin{align}
 &\label{BAE'-1}
 \frac{\xi'_\alpha}{\xi_\alpha}
 +\frac{\beta\mu'_j}{\mu_j+\frac{\rm i}2}
 +\frac{\mu'_j}{\mu_j}
 +\frac{(2N+1+\beta)\mu'_j}{\mu_j-\frac{\rm i}2}
 =\frac{\mu'_j}{\mu_j+{\rm i}p'}
 \nonumber\\ &
 +\frac{\mu'_j}{\mu_j+{\rm i}q'}
 +\sum_l\frac{\mu'_j+\mu'_l}{\mu_j+\mu_l}
 +\sum_l\frac{\mu'_j+\mu'_l}{\mu_j+\mu_l-{\rm i}},
\end{align}
where $\mu'_j=\partial_\alpha \mu_j$ and
$\xi'_\alpha=\partial_\alpha \xi_\alpha$. From Eq.(\ref{BAE'-1}), we
can see that when $N$ is very large, $\mu'_j$ are very small. Up to
the leading order the above equations can be written as
\begin{align}
 \label{BAE'b}
 &\mu'_j\approx
 \frac{1}{N}
 \frac{\xi'_\alpha}{\xi_\alpha}
 \Big[
  \frac1N\sum_l\Big(\frac{2}{\mu_j+\mu_l}
 +\frac{2}{\mu_j+\mu_l-{\rm i}}\Big)
 -\frac{2}{\mu_j-\frac{\rm i}2}
 \Big]^{-1},
\end{align}
which indicate that $\mu_j'$ are in the order of $1/N$. This means
that for two different $\alpha$ the maximum difference of $\mu_j$
for the corresponding state is $\Delta \mu_j \lesssim 1/N$. The
numerical results in TABLE \ref{tablered} for $N=10$ also support
this indication. When the lattice number $N$ is very large, the
boundary fields only affect the spins close to the ends. Indeed, the
DMRG results showed that the correlations of the spin at the ends
and the ones in the bulk are almost independent of the angle
$\alpha$. ~\cite{zhuo} To show the contribution of $\alpha$ to the
energy, we divide the eigen energy $E$ of the Hamiltonian (\ref{H})
into two parts
\begin{align}
 \label{DE}
 E(p,q_0,\alpha)=E(p,q_0,0)+\Delta E(p,q_0,\alpha),
\end{align}
$E(p,q_0,0)$ is the energy for $\alpha=0$  and $\Delta E$ is the
contribution of nonzero $\alpha$.
\begin{figure}[t]
 \includegraphics[width=\linewidth]{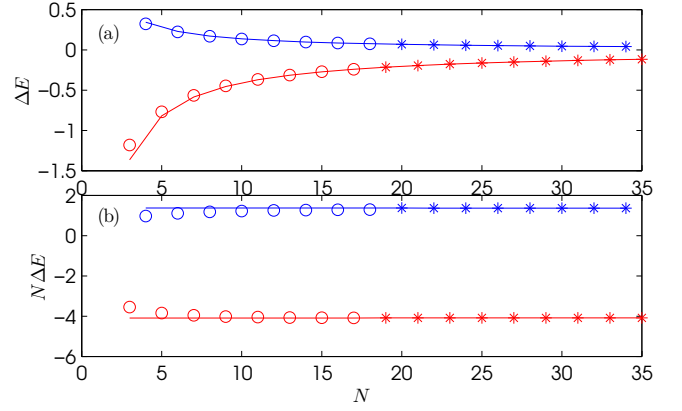}
\caption{\label{dirc}
 (color online) Ground state $\Delta E$ for $\alpha=\pi/2$, $p=0.6$ and $q_0=0.6$.
 The circles and stars are results from exact diagonalization and matrix product states respectively.
 The lines are the fittings with the formula $\Delta E=b_0+b_1/N$.
 The blue/red lines and points are for even/odd $N$, respectively.
 For even $N$, the points are fitted from the data of $N\geq16$ with $b_0=-0.0002$ and $b_1=1.3710$;
 while for odd $N$ the points are fitted from the data of $N\geq17$ with $b_0=-0.0001$ and $b_1=-4.0826$.}
\end{figure}
For a large boundary field angel $\alpha=\pi/2$, we plot $\Delta E$
in FIG. \ref{dirc}. FIG. \ref{dirc} (a) gives the $\Delta E$ and
FIG. \ref{dirc} (b) gives the $N\Delta E$. The numerical fitting
shows that $\Delta E$ is of the order $1/N$. For the contribution of
the boundary fields to the energy is of order $O(N^0)$, in the
thermodynamic limit we can omit the contribution of $\alpha$ up to
the leading order. In fact, the two boundaries are decoupled from
each other completely when $N\to\infty$.

\section{Conclusion}
\label{SecC}

In conclusion, we numerically verified the completeness and the
correctness of the off-diagonal Bethe ansatz solutions for the
Heisenberg spin chain model with arbitrary boundary fields. The
one-to-one correspondence between the Bethe roots of the unparallel
boundary field case and those of the parallel boundary field case is
established based on the reduction process. This correspondence
allows us to study the physical effect of the relative angle
$\alpha$ by a proper perturbation approach, since the Bethe root
distribution of the parallel boundary field case is already well
known. It is shown that the contribution of $\alpha$ to the energy
is of order $N^{-1}$, which is negligible small in the thermodynamic
limit $N\to\infty$. We remark that such correspondence and
perturbation procedure are also suitable for other models solved via
the off-diagonal Bethe ansatz.

The financial support from  the National Natural Science Foundation
of China (Grant Nos.11174335, 11075126, 11031005, 11375141,
11374334), the National Program for Basic Research of MOST (973
project under grant No.2011CB921700) and the State Education
Ministry of China (Grant No. 20116101110017 and SRF for ROCS) are
gratefully acknowledged.

\end{document}